# Epitaxial Recovery of $\beta$-Ga$_2$O$_3$ after High Dose Implantation

Tianhai Luo[1], Katie Gann[1,2], Cameron Gorsak[1], Hari P. Nair[1], R. B. van Dover[1], and Michael O. Thompson[1]

[1]Department of Materials Science and Engineering, Cornell University, Ithaca, NY 14853, USA

[2]U.S. Naval Research Laboratory, Washington, DC 20375, USA

**Abstract**

As an ultrawide bandgap semiconductor, β-Ga$_2$O$_3$ has been attractive for its strong tolerance to irradiation damage and high n-type conductivity through ion implantation. Homoepitaxial (010) β-Ga$_2$O$_3$ films grown by MOCVD were implanted with Ge to study the post-implantation damage and lattice recovery after thermal annealing. Box profiles of 100 or 50 nm at concentration of 5×10$^{19}$ or 3×10$^{19}$ cm$^{-3}$ were formed, with maximum displacement per atom (DPA) of 1.2 or 2.0. Lattice recovery was investigated using X-ray diffraction (XRD) for anneals from 100°C to 1050°C. A γ-phase related peak was observed for all implant conditions. All samples showed strain relaxation of β-phase peak at temperature below 500 °C, with no significant change for the γ-phase related peak. For lower damage implants, films recovered fully to epitaxial β-phase after sequential annealing to 900 °C. For the higher damage implant, the γ-phase associated peak annealed out with increasing temperature, but a new diffraction peak formed at slightly smaller lattice spacing; full recovery of the lattice was not observed until annealing at 1050 °C. The newly formed diffraction peak is identified as β-(20$\bar{4}$), β-(512), or β-(71$\bar{2}$), each potentially arising from the conversion of γ-phase to β-phase via a common oxygen sub-lattice.

**Introduction**

The ultrawide bandgap of β-Ga$_2$O$_3$ and its availability for scale-up substrate manufacture using melt-grown crystals has demonstrated great potential for future power electronics[1]. Its advantage extends to the superior tolerance to radiation damage[2], benefiting application in harsh environment and functionalization through dopant implantation. Despite a high tolerance to radiation damage and amorphization is not observed at room temperature[2], under high dose implantation a transformation to the γ-phase has been observed[3–6]. So far, these implantation-induced transformations have been well acknowledged but there is only limited data relating lattice damage and recovery to the implantation dose, concentration and damage profile. In this work, we studied the lattice damage of β-Ga$_2$O$_3$ subjected to various implantation sequences, and its recovery after thermal annealing as a function of anneal temperature.

**Experimental**

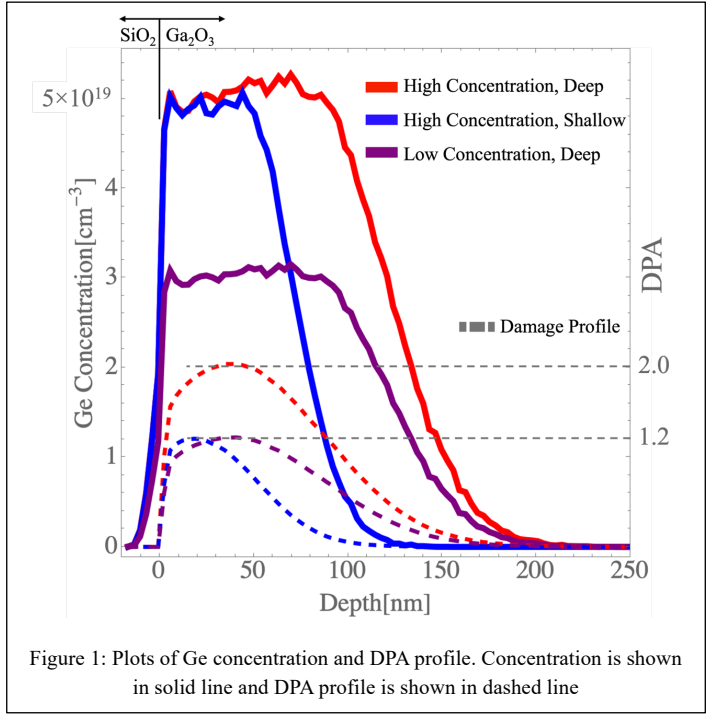

Figure 1: Plots of Ge concentration and DPA profile. Concentration is shown in solid line and DPA profile is shown in dashed line

Fe-doped (010) β-Ga$_2$O$_3$ substrates (24×25 mm) were obtained from Novel Crystal Technology and a 500 nm unintentionally doped (UID) buffer layer was grown by MOCVD using conditions reported previously[7]. Prior to implantation, a 20 nm SiO$_2$ cap was deposited by ALD to ensure constant doping concentration in Ga$_2$O$_3$ at the interface. A single 24×25 substrate was diced into individual samples for the three Ge implant conditions to ensure nominally identical samples for measurements. Using four energies (Table 1), box-like concentration profiles were formed to 50 nm and 100 nm at target concentrations of 3 and 5×10$^{19}$ cm$^{-3}$. The SiO$_2$ cap was removed using a 60 second etch in 6:1 buffered oxide etchant (BOE) before annealing. Solid lines in figure 1 show the estimated distribution of Ge dopants in depth as determined by the Stopping Range of Ion in Matter (SRIM).[8]

The three implant conditions were designed to cover a range of concentrations and depths commonly explored for Si implantation[9,10], and a range of implant damage conditions. The dashed lines in figure 1 shows the calculated damage profiles (displacements per atom or DPA) for the three implant schedules. Samples will be identified by the nominal implant concentration and depth; H/D refers to high concentration (5×10$^{19}$) and deep (100 nm), H/S to high concentration (5×10$^{19}$) and shallow (50 nm), and L/D to low concentration (3×10$^{19}$) and deep (100 nm). Samples H/D and H/S differ in the box depth at similar Ge concentration. Sample L/D extends the implant to 100 nm with a dose approximately the same as the H/S sample. The maximum DPA for H/S and L/D samples are comparable near 1.2 and will be collectively referred to as low-damage conditions, while the H/D sample has a greater damage of maximal DPA at 2.0 and is referred to as the high-damage implant. These samples permit comparison of lattice damage due to ion implantation as a function of dopant concentration, depth, dose, and damage (DPA). After implantation, the substrates were diced into 5×4 mm samples for annealing.

| Identifier | Box depth (nm) | Target box concentration (×$10^{19}$ cm$^{-3}$) | Total dose ($10^{14}$ cm$^{-2}$) | Box dose ($10^{14}$ cm$^{-2}$) | Max DPA | Implant protocol (keV energy / $10^{13}$ cm$^{-2}$ dose) |
|---|---|---|---|---|---|---|
| High concentration, Deep (H/D) | 100 | 5.0 | 6.64 | 5 | 2.0 | 25/3.1; 60/5.8; 120/11; 260/46.5 |
| High concentration, Shallow (H/S) | 50 | 5.0 | 3.87 | 2.4 | 1.2 | 25/3.1; 60/5.6; 70/1.0; 160/29 |
| Low concentration, Deep (L/D) | 100 | 3.0 | 3.98 | 3 | 1.2 | 25/1.9; 60/3.5; 120/6.6; 260/27.8 |

Table 1. Details for Ge implant conditions

Figure 2 shows XRD spectra of as-implanted samples in the region of the (020) $\beta$-phase peak. In addition to the substrate peak and strain (damage) broadening at 61°, an additional peak near 63.5° was observed for all implant conditions. Previous studies on implantation in β-Ga$_2$O$_3$ have identified this peak as arising from $\gamma$-phase inclusions[4,6,11,12] induced by the radiation damage above a critical dose. The $\gamma$-phase inclusions are highly strained within the local crystal environment, resulting in the $\gamma$-phase related peak shifting from the expected 63.9° angle to around 63.5°. The location and shape of the $\gamma$-phase related peak is comparable for both low-damage samples (L/D and H/S). However, the high damage sample (H/D) exhibits a sharper peak at a slightly higher angle (closer to the expected 440 $\gamma$-phase reflection). In addition to the $\gamma$-phase peak, a high angle-shoulder on the (020) β-peak was also observed. This strain shoulder has also been reported previously for Si implantation and was attributed to compressive strain in as-implanted β-Ga$_2$O$_3$[10,12–15].

These $\gamma$-phase related peaks are not observed for Si implants with damage at or below 1.5 DPA, appearing only for samples implanted to box profile for 150 nm at concentration of $10^{20}$ cm$^{-3}$ with a maximal DPA of 2.0.[10] As DPA is calculated from simulated vacancy densities formed during implantation, it is expected to be a good metric of damage independent of the dopant species and hence the $\gamma$-phase peak was not expected in the low-damage samples of this experiment. The presence of this damage peak implies that Ge is either much more effective than Si in generating the critical defects leading to $\gamma$-phase inclusions, or that the damage is strongly influenced by dose rate[16] or implant temperature[17]; these parameters were not specifically controlled in these experiments. For Si implants, it has also been observed that annealing of samples with these $\gamma$-phase related defects becomes significantly more difficult relative to samples with no $\gamma$-phase signature in XRD.[10]

To study the lattice thermal recovery, samples were annealed at elevated temperature sequentially under N$_2$. Figure 4 shows XRD spectra measured after each temperature. Figure 4a shows the low temperature initial recovery of sample L/D annealed from 150°C to 500°C. Thermal recovery of all samples behaves similarly at temperatures below 500 °C and only the case of L/D is shown. Figures 4b shows the thermal recovery of sample L/D from 500°C to 1050°C. The thermal recovery of sample H/S is similar to L/D in this range of temperature. Figure 4c shows thermal recovery of the higher damage H/D sample from 500 °C to 1050 °C.

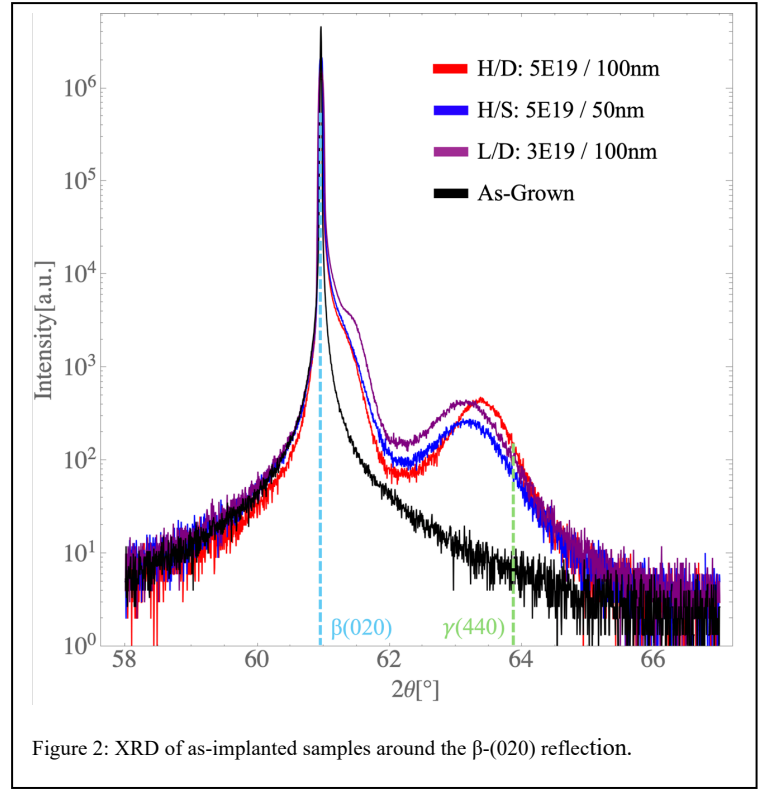

Figure 2: XRD of as-implanted samples around the β-(020) reflection.

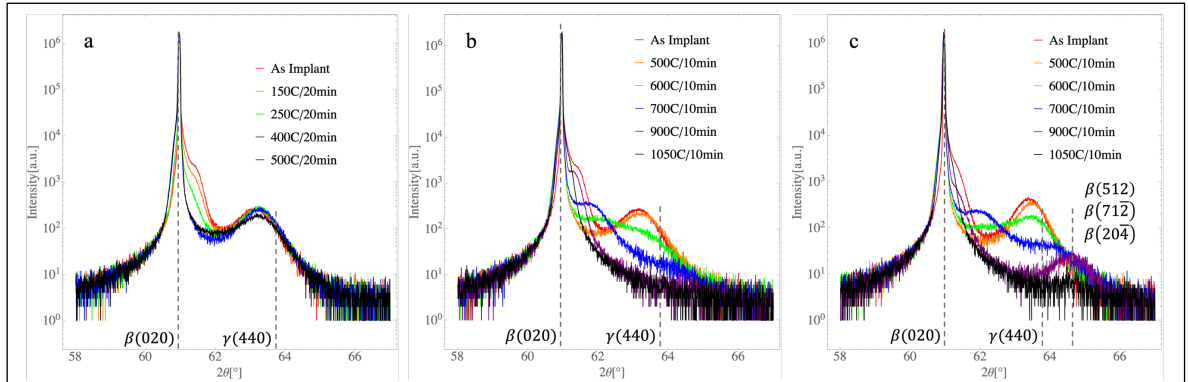

Figure 4: XRD spectra showing lattice recovery for (a) sample L/D from 150°C to 500°C, (b) sample L/D from 500°C to 1050°C, (c) sample H/D from 500°C to 1050°C.

After lower temperature anneals, figure 4a shows strain relaxation within the β phase matrix as a reduction of the high-angle β-phase shoulder, disappearing entirely for anneal temperature ≥400 °C. This is followed by the formation of a shoulder on the low angle side at 500°C. For temperature below 500°C, no change in the γ-phase peak was observed. For annealing temperatures above 500°C, the γ-phase inclusions begin to disappear with the peak decreasing. Figure 4b shows XRD spectra of an L/D sample sequentially annealed from 500°C to 1050°C, with reduction in the γ-phase related peak beginning at 500°C. For temperatures above 900°C, there is no observable peak remaining in the low damage samples (the high damage H/D samples are discussed separately below). This conversion of γ-phase damage to β-phase agrees well with previous reports following Si implantation[3,18–21]. Concurrent with the loss of the γ-phase related peak, a high-angle shoulder on the (020) β-peak develops. Beginning to develop above 600°C, it develops into a strong and wide shoulder at 700°C before ultimately collapsing to a very high intensity but narrow shoulder above 900°C. Similar behavior was observed for both low-damage samples. See SI for XRD spectra from complete temperature steps.

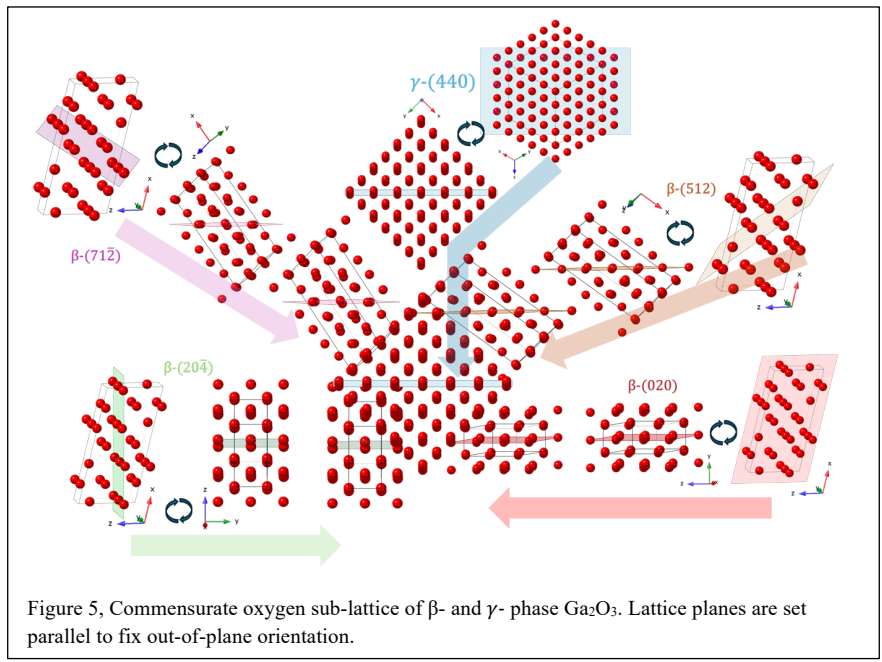

Figure 5, Commensurate oxygen sub-lattice of β- and γ- phase Ga$_2$O$_3$. Lattice planes are set parallel to fix out-of-plane orientation.

For the high damage level, sample H/D exhibits different lattice recovery behavior. Starting with a γ-phase related peak closer to standard (440) γ-phase reflection at 63.9°, H/D sample is indicated to form larger γ-phase inclusions with more "γ-like" behavior with the higher damage from implantation. For low temperature anneals below 500°C, it behaves similarly to the low-damage Ge implants, with annealing of the β-phase shoulder but no change in the γ-phase peak. The behavior at higher temperatures is significantly different. Figure 4c shows the XRD spectra of this high-damage sample sequentially annealed from 500°C to 1050°C. For anneals from 500°C to 900°C, the γ-phase related peak decreases and the shoulder on the higher angle of (020) β-peak develops.

Based on previous studies, we believe that γ-phase itself cannot be retained at 900°C[21,22] with full conversion to β-phase expected. We believe this peak represents conversion of γ-phase like regions to different orientations of β-phase other than (010), that are also related through the oxygen sub-lattice. Theories on the mechanism of β to γ phase transition have been proposed by various authors including Wouters[23], Azarov[19] and Hwang[5,6]. We believe the γ-phase regions of the as-implanted structure share a common oxygen sub-lattice with the initial β-phase[24], stabilized with small strains to match the oxygen sub-lattice. Conversions between the β- and γ-phase are postulated to be based on Ga cations hopping between respective Wycoff sites within the lattice. Specifically, the oxygen sub-lattice of β-phase viewed from [010] would aligns with that of γ-phase viewed from [110][23]. This explains why γ-(440) reflections related peak is observed for epitaxial (010) β-Ga$_2$O$_3$ when subjected to implant damage. For the low damage samples (H/S and L/D), γ-phase inclusions could fully recover to β-phase epitaxy at 900°C. For the high-damage H/D sample, larger γ-phase inclusions are formed during implantation and would recover to β-phase with other orientations sharing the commensurate oxygen sub-lattice. Based on this theory, we identified three candidate reflections for the new formed peak: β-(20$\bar{4}$), β-(512), or β-(71$\bar{2}$).

Figure 5 shows the commensurate oxygen sub-lattice of β- and γ- phase. β-(20$\bar{4}$), β-(512), β-(71$\bar{2}$), β-(020) and γ-(440) planes are demonstrated in standard unit cell, followed by a rotation process to align out-of-plane orientation. This rotation process allowed these planes to satisfy the Bragg condition to be observed in symmetrical $2\theta$-ω XRD scan. Post-rotation unit cells were merged in the center, demonstrating a good alignment of oxygen sub-lattice. The merged-lattice is viewed from γ-[00$\bar{1}$] with a slight tilt to show that these planes are parallel and therefore all of them could be observed in a $2\theta$-ω symmetrical scan aligned to any one of the planes. The super unit cell with well-aligned oxygen sub-lattice shows that the proposed planes could be observed in symmetrical XRD scans sharing the oxygen sub-lattice from bulk 010 orientation.

**Conclusion**

In conclusion, irradiation damage and thermal recovery was studied by XRD on Ge implanted β-Ga$_2$O$_3$ with various profiles. The location of γ-phase damage peak is shown to be related to implant damage, with higher damage resulting γ-phase related peak closer to theoretical angle. Initial lattice recovery was identified as strain relaxation, occurring at temperature ≤500 °C. Annealing of γ-phase damage begins at 500 °C and finishes at 900 °C. For higher damage sample, one additional peak at higher angle remained until annealing at 1050 °C. This extra peak is identified as arising from 20$\bar{4}$, 512, and 71$\bar{2}$ planes of β-phase. These planes could satisfy Bragg condition with the bulk 010 β-phase epitaxy while maintaining the common oxygen sublattice.